# Quantitative evaluation of nuclear quantum effects on the phase transitions in BaTiO$_3$ using large-scale molecular dynamics simulations based on machine learning potentials

Kansei Kanayama* and Kazuaki Toyoura

*Department of Materials Science and Engineering, Kyoto University, Kyoto 606-8501, Japan*

*kanayama.kansei.67z@st.kyoto-u.ac.jp

**Abstract**

The machine learning potential (MLP) based molecular dynamics (MD) method was applied for constructing the pressure-temperature phase diagram in the barium titanate (BaTiO$_3$) crystals. The nuclear quantum effects (NQEs) on the phase transitions were quantitatively evaluated from the difference in the phase transition pressures between the NQEs-incorporated and classical simulations. In this study, the quantum thermal bath (QTB) method was used for incorporating the NQEs. The constructed phase diagrams verified that the NQEs lower the phase transition temperatures and pressures. The NQEs are more significant at lower temperatures but cannot be ignored even at room temperature. The phase diagram in the QTB-based MLPMD is in good agreement with those of the previous studies based on dielectric measurements and path-integral based simulations. In addition, this study clarified that the large cell size (a 16×16×16 or larger cell) and friction coefficient ($\geq$ 15 THz) are required for accurately reproducing the phase transitions during the QTB-MD simulations.

# I. Introduction

Barium titanate, BaTiO$_3$, is a typical ferroelectric material. It has sequential phase transitions from the paraelectric cubic phase to the ferroelectric tetragonal, orthogonal, and rhombohedral phases with decreasing temperature or pressure [1–6]. While every Ti ion is positioned at the center of the octahedron formed by six O ions in the cubic phase, it is displaced from the center along with [001], [011], and [111] directions in the tetragonal, orthogonal, and rhombohedral phases, respectively.

Many experiments and theoretical calculations have been performed to determine the phase transition temperatures or pressures in BaTiO$_3$. The transition temperatures of the rhombohedral-orthogonal (*R-O*), orthogonal-tetragonal (*O-T*), and tetragonal-cubic (*T-C*) phase transitions are reported to be 183, 273, and 393 K at ambient pressure, respectively, which were determined through X-ray diffraction measurements [2]. A previous study using Raman spectroscopy reported that the transition pressure of the *T-C* transition is around 2 GPa at room temperature [7]. Several dielectric measurements [8,9] clarified that the *T-C* transition temperature linearly decreases with a decrease in pressure above room temperature. Ishidate et al. [10] determined a pressure-temperature phase diagram by dielectric measurements in the range of 4–400 K and 0–8 GPa. In the phase diagram, the phase transition temperature $T_c$ and pressure $P_c$ have a relation of $T_c \propto (P_c - P_c(0\ \mathrm{K}))^{1/2}$ at low temperatures, where $P_c(0\ \mathrm{K})$ is the transition pressure at 0 K.

In theoretical studies, the transition temperatures and pressures have been investigated based on the first-principles calculations [11–16]. Ehsan et al. [11] reported the *T-C* transition temperature to be 455 K by the self-consistent phonon calculations [17–19] based on the PBEsol exchange-correlation functional. Zhong et al. [12,13] made a pressure-temperature phase diagram based on the classical Monte Carlo (MC) simulations using an effective Hamiltonian based on the LDA. Inigues et al. [14] and Geneste et al. [15] performed the path integral MC and MD simulations [20–22], respectively, for incorporating nuclear quantum effects (NQEs), where the same effective Hamiltonian constructed by Zhong et al. were employed. The pressure-temperature phase diagram obtained by the path integral MC [14] was significantly different from the one obtained by the classical MC, indicating the importance of the NQEs at low temperatures below room temperature. The MD simulations with the quantum thermal bath (QTB) [23–25], which is another method for incorporating the NQEs, were also applied to determine the phase transition temperature under ordinary pressure in BaTiO$_3$ [16].

Although these theoretical studies suggest the importance of the NQEs on the phase transitions in BaTiO$_3$, they basically employed the effective Hamiltonian in which the degrees of freedom (DOFs) in the system were reduced to the several phonon modes corresponding to the TO and acoustic modes of the cubic structure. We have therefore revisited the quantitative evaluation of the NQEs on the pressure-temperature phase diagram in the present study, where all the DOFs in the system were taken into consideration. For incorporating the NQEs into MD simulations, we employed the QTB method because of the low computational cost comparable to the conventional MD with the classical thermal bath (CTB-MD), which was demonstrated in our previous study on the phase transition in $CdTiO_3$ [26]. Brieuc et al. [16] reported that the QTB-MD simulations with a large friction coefficient are able to reproduce the NQEs in the path integral MD requiring much higher computational costs.

Machine learning potentials (MLPs) were here used to reduce the computational costs of MD simulations with a large amount of DOFs. Once appropriately constructed using first-principles data for training, the MLPs can reproduce the total energy, atomic forces, and stress of a given system with an accuracy comparable to that of the first-principles calculations [27–30]. We constructed the MLPs through the on-the-fly scheme during the MD simulations [30]. We carefully checked the accuracy of the constructed MLPs for $BaTiO_3$ and the dependencies of the phase transitions on the friction coefficient and cell size. Then, the phase diagram as a function of temperature and pressure was determined by the QTB-MD and CTB-MD to quantitatively evaluate the NQEs on the phase transitions. The obtained phase diagram was finally compared with those reported in the literature, i.e., the path integral MC study [14] and the dielectric measurement study [10].

## II. Computational method
### A. Quantum thermal bath method

QTB-MD is based on the Langevin MD [31], in which a motion of each DOF is described by the Langevin equation [23–25].

$$m_i \frac{d^2 u_{i\alpha}}{dt^2} = -\frac{\partial U}{\partial u_{i\alpha}} - m_i \gamma \frac{du_{i\alpha}}{dt} + R_{i\alpha}(t), \quad (1)$$

where $m_i$ and $u_{i\alpha}$ are the mass and three-dimensional cartesian coordinate ($\alpha$ = 1, 2, 3) of the $i$th atom, respectively, and $t$, $U$, and $\gamma$ are the time, the potential energy, and the friction coefficient, respectively. In a Langevin MD simulation, the temperature is controlled by the friction and random forces corresponding to the second and third terms in the right-hand side of Eq. (1). At every MD step, $\partial U/\partial u_{i\alpha}$ is calculated, the random force is generated, and $u_{i\alpha}$ and $du_{i\alpha}/dt$ are updated according to Eq. (1). In a conventional Langevin MD with the CTB, a Gaussian random force is given to reproduce the energy equipartition theory. Specifically, the time correlation function of $R_{i\alpha}(t)$ has no time-dependency, represented as $2m_i\gamma k_B T\delta(t)$, where $k_B$, $T$, and $\delta(t)$ are the Boltzmann constant, the temperature, and the delta function, respectively. The mean kinetic energy per DOF in a CTB-MD coincides with $1/2\ k_B T$.

In a QTB-MD simulation, a time-correlated random force is given in contrast to the time-uncorrelated random force in a CTB-MD simulation. Although each DOF is treated as a classical particle, the NQEs are incorporated as the kinetic energy through the random force. The power spectral density $I_R(\omega)$ of the time-correlated random force is expressed as

$$I_R(\omega) = 2m_i\gamma \left( \frac{1}{2}\hbar|\omega| + \hbar|\omega| \frac{1}{\exp\left(\frac{\hbar|\omega|}{k_B T}\right) - 1} \right), \quad (2)$$

where $\hbar$ and $\omega$ are the Dirac constant and the angular frequency, respectively. Under the harmonic approximation, the mean energy coincides with the quantum harmonic energy of the lattice vibration, which is composed of the zero-point energy and the phonon excitation term.

### B. Computational conditions

In this study, the on-the-fly scheme [30] was used for constructing the MLPs for BaTiO$_3$, which is implemented in the VASP code [32–34]. In this scheme, structure datasets for constructing MLPs are updated in a single first-principles MD (FPMD) simulation. When the predicted error of the force field on a newly sampled structure is smaller than a given threshold at each MD step, the position and velocity of each atom in the system are updated according to the constructed MLPs without the first-principles calculations. Otherwise, the first-principles calculation is performed for the sampled structure, and the MLPs are reconstructed using the new training datasets including the sampled structure.

The MLPs were constructed by the on-the-fly scheme at each pressure $P$, $P = 0$, 1, …, 17 GPa. The temperature in each FPMD simulation was linearly increased from 100 to 400 K with the CTB for learning structures at different phases. The simulation time and step size were 60 ps and 1 fs, respectively, and the friction coefficient was 20 THz. The initial structure of each FPMD simulation was a 4×4×4 pseudo-cubic cell with the rhombohedral structure, which was optimized at each pressure. A projector augmented wave (PAW) method [35] was used, in which the cut-off energy and $k$-point sampling were 500 eV and the Γ-point only, respectively. The generalized gradient approximation with Perder-Burke-Ernzerhof parametrization (PBE_GGA) [36] was used as the exchange-correlation functional, where electrons in 5s, 5p, and 6s for Ba, 3p, 3d, and 4s for Ti, and 2s and 2p for O were treated explicitly as valence electrons.

MD simulations based on the MLPs (MLPMD) were performed at each temperature $T$ and pressure $P$, $T = 100, 150, 200, 250, 300$ K and $P = 0, 1, …, 17$ GPa, with the QTB or CTB. In this study, the QTB code for generating a time-correlated random force was implemented into the VASP code. As in the FPMD simulation, the initial structure of each MLPMD simulation was a pseudo-cubic structure of the rhombohedral phase, and the step size and the friction coefficient were 1 fs and 20 THz, respectively. For verifying the cell-size effects on the phase transitions, QTB- and CTB-MLPMD simulations were performed at different cell sizes: supercells consisting of 12×12×12, 16×16×16, and 20×20×20 pseudo-cubic cells. The simulation time was 60 ps including 40 ps for thermal equilibrium. In a QTB-MD simulation, the PSD in Eq. (2) diverges proportionally to the angular frequency $\omega$ for high $\omega$. Thus, the maximum of $\omega$ should be specified for avoiding the divergence of the PSD, which was set as $50\pi$ THz.

## III. Results and discussion
### A. Verification of computational conditions
#### 1. Machine learning potentials

The accuracy of the constructed MLPs was verified against the first-principles (FP) calculations based on the PBE_GGA. The QTB- and CTB-MLPMD simulations at $4\times4\times4$ and $5\times5\times5$ supercells were performed under the conditions of $T = 100, 300$ K and $P = 0, 5, 10, 15$ GPa. A hundred structures were sampled at every 500 fs after the thermal equilibrium steps for 10 ps under each condition. The total energy, atomic forces, and stress of each structure were then computed by the FP calculations to quantitatively evaluate the deviation of the constructed MLPs from the FP calculations.

Figure 1 shows the total energies, atomic forces at titanium ions, and stresses computed in the MLPs as a function of those computed in the FP calculations. The plotted data in Fig. 1 corresponds to the structures sampled during the MLPMD at 10 GPa. The results corresponding to 0, 5, and 15 GPa are shown in Figs. S1-S3 in Supplemental Materials. The total energies, atomic forces, and stresses ideally have linear relations through the origin with a slope of unity between the MLPs and FP calculations. The root mean square errors (RMSEs) of the atomic forces and stresses predicted by the MLPs were 51 meV/Å and 0.029 GPa at a $4\times4\times4$ supercell, and 77 meV/Å and 0.047 GPa at a $5\times5\times5$ supercell, respectively. Thus, the constructed MLPs reproduce the atomic forces and stresses computed in the FP calculations within an acceptable error. For the total energies, the RMSEs at $4\times4\times4$ and $5\times5\times5$ supercells were 0.12 and 1.90 meV/atom, respectively. Although the MLPs reproduced the total energies with a first-principles accuracy at both supercells, a systematic error can be seen at a $5\times5\times5$ supercell. The magnitudes of the systematic errors are comparable between the different temperatures. The mean difference in the total energies between the MLPs and FP calculations was 1.89 eV/atom at a $5\times5\times5$ supercell, while the RMSE without the systematic error was 0.14 meV/atom comparable to that at the $4\times4\times4$ supercell. Since the systematic errors are expected to have negligible influence on the estimated phase transition temperatures and pressures, we used the MLPs constructed at the $4\times4\times4$ supercell for the MD simulations at larger supercells in the present study, i.e., $12\times12\times12$, $16\times16\times16$, and $20\times20\times20$ supercells.

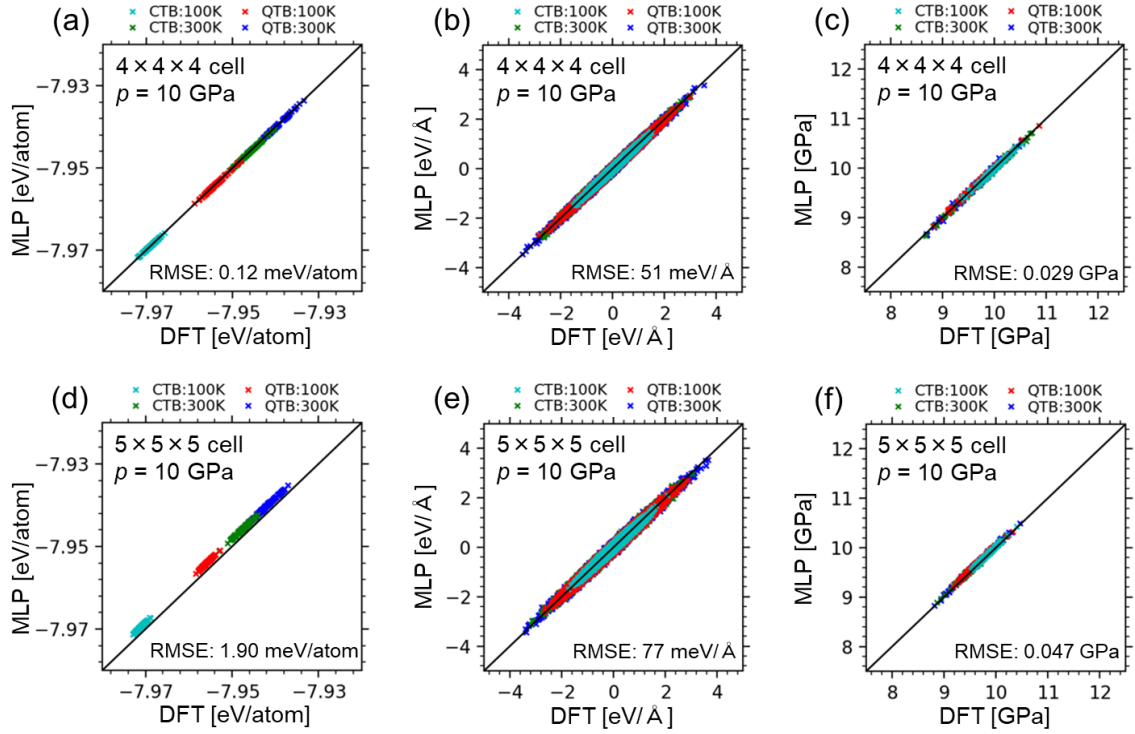

Fig. 1. (a,d) Total energies, (b,e) atomic forces on titanium ions, and (c,f) stresses computed in the constructed MLPs as a function of those in first-principles calculations. The data corresponds to the structures sampled during the MLPMD simulations at 10 GPa with (a-c) 4×4×4 and (d-f) 5×5×5 pseudo-cubic structures.

## 2. Friction coefficient in the QTB method

Brieuc et al. [16] reported that the rhombohedral and/or orthogonal phases were missed in QTB-MD simulations with a low friction coefficient $\gamma$. For determining an appropriate friction coefficient, the QTB-MLPMD simulations at 0 GPa were performed under the conditions of $T$ = 100, 150, 200 K and $\gamma$ = 5, 10, 15, 20 THz. A 16×16×16 supercell was employed in these simulations, in which the pseudo-cubic rhombohedral phase was used as the initial structure. Figure 2 shows the profiles of each lattice parameter during the QTB-MLPMD simulations as a function of time. In the pseudo-cubic structures, the lattice parameters $a$, $b$, and $c$ have the following relations: $a = b = c$ in the cubic and rhombohedral phases, $a = b < c$ in the tetragonal phase, and $a < b = c$ in the orthogonal phase. In the early stage of these simulations, the lattice parameters have the relation of $a = b = c$ reflecting the initial rhombohedral phase regardless of the simulation conditions. At 200 K, the relation between the lattice parameters changes from $a = b = c$ to $a < b = c$ after 40 ps with any friction coefficient in the range of $5 \leq \gamma \leq 20$ THz. This change in the lattice parameters corresponds to the structural change from the rhombohedral to the orthogonal, indicating that the orthogonal phase is stable at 200 K. Note that the stable phase depends on the friction coefficient at 100 and 150 K. With the low friction coefficients ($\gamma$ = 5 THz at 100 K and $\gamma$ = 5, 10 THz at 150 K), the lattice parameters show similar profiles to those at 200 K, corresponding to the structural change from the rhombohedral to the orthogonal. By contrast, with the high friction coefficients ($\gamma$ = 10, 15, 20 THz at 100 K and $\gamma$ = 15, 20 THz at 150 K), the lattice parameters had always remained the relation of $a = b = c$ during the simulations, meaning that the rhombohedral phase is stable under these conditions. Considering the experimental phase transition temperature from the rhombohedral to the octahedral, $T_{R-O}$ = 183 K [2], the friction coefficient should be set as 15 THz or larger in this system. A possible reason for the lowering of the phase transition temperature is the zero-point energy leakage from high-frequency vibrational modes to low-frequency ones, which is a well-known problem in QTB-MD simulations [16,23,37–39]. $\gamma$ = 20 THz was adopted in the present study to avoid significant leakage.

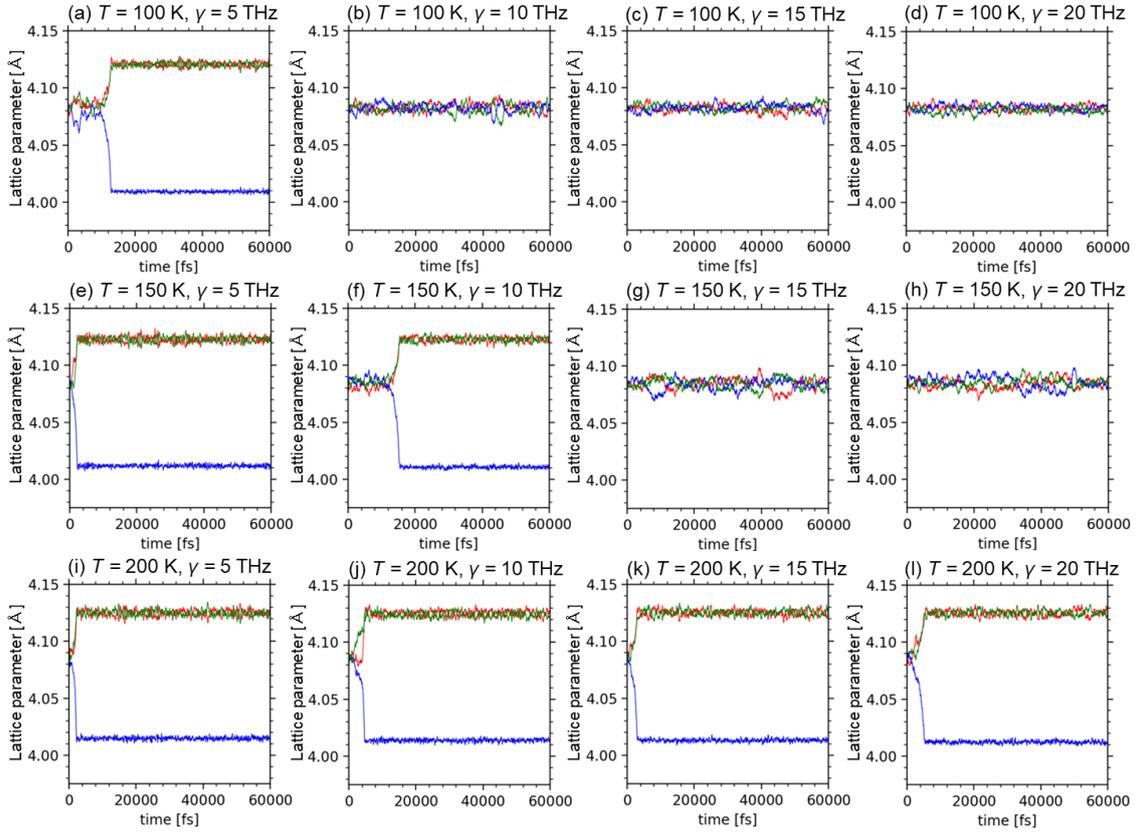

Fig. 2. Lattice parameter profiles during the QTB-MLPMD simulations as a function of time. The MD simulations at 0 GPa were performed under the conditions of (a-d) 100 K, (e-h) 150 K, and (i-l) 200 K with the friction coefficient of (a,e,i) 5 THz, (b,f,j) 10 THz, (c,g,k) 15 THz, and (d,h,l) 20 THz.

## 3. Cell-size dependency of the phase transitions

To determine an appropriate cell size for analyzing the phase transitions, QTB-MLPMD simulations at 150 K were performed in 12×12×12, 16×16×16, and 20×20×20 supercells under a constant pressure in the range of 0–12 GPa. The two structural parameters were monitored for determining the phase transition pressures: the lattice parameters ($a$, $b$, $c$) and the relative ionic displacement of Ti ions against O ions, $\Delta = (\Delta_a, \Delta_b, \Delta_c) = |\boldsymbol{g}_{Ti} - \boldsymbol{g}_O|$, where $\boldsymbol{g}_X$ is the center of $X$ ions in a pseudo-cubic cell. The lattice parameters with each condition were estimated from the distributions of lattice parameters during the simulation except the thermal equilibrium steps as shown in Fig. 3. Specifically, the distribution of lattice parameters was first obtained by accumulating standard normal distributions, represented as

$$f(x) = \frac{1}{3N} \sum_i \frac{1}{\sqrt{2\pi\sigma^2}} \left( \exp\left(-\frac{(x-a_i)^2}{2\sigma^2}\right) + \exp\left(-\frac{(x-b_i)^2}{2\sigma^2}\right) + \exp\left(-\frac{(x-c_i)^2}{2\sigma^2}\right) \right), \quad (3)$$

where $a_i$, $b_i$, and $c_i$ are the monitored lattice parameters at time step $i = 1, \ldots, N$ without the thermal equilibrium steps, and $\sigma$ was set to 0.0002 Å. The accumulated distribution of lattice parameters was then fitted by three normal distributions, represented as

$$f(x) = \frac{1}{3} \sum_{j=1,2,3} \frac{1}{\sqrt{2\pi\sigma_j^2}} \exp\left(-\frac{(x-\mu_j)^2}{2\sigma_j^2}\right) \quad (4)$$

where the fitting parameters are $\mu_1$, $\mu_2$, $\mu_3$, $\sigma_1$, $\sigma_2$, and $\sigma_3$, and the former three parameters correspond to the lattice parameters $a$, $b$, and $c$. The relative ionic displacements of Ti ions against O ions were estimated in the same manner, where $\sigma$ in Eq. (3) was set to 0.0002 Å.

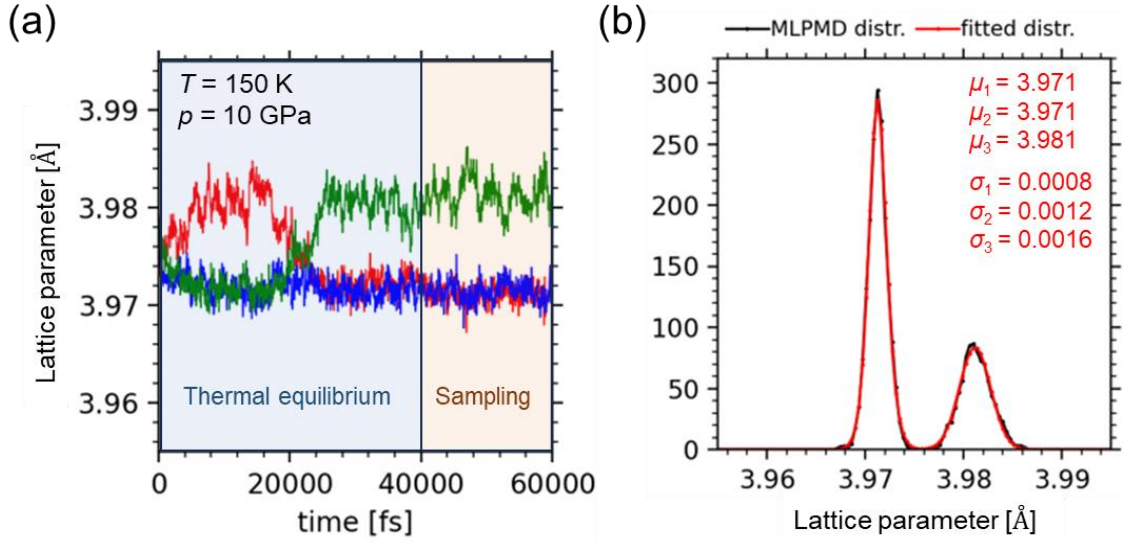

Fig. 3. (a) Lattice parameter profiles during the QTB-MLPMD simulation at 150 K and 10 GPa in a 16×16×16 supercell. (b) Lattice parameter distribution during the QTB-MLPMD (black line) generated by accumulating standard normal distributions in Eq. (3). The red line corresponds to the fitted distribution, where the lattice parameters were estimated as $\mu_1$, $\mu_2$, and $\mu_3$ in Eq. (4).

Figures 4 and 5 show the estimated lattice parameters and the relative ionic displacements of Ti ions against O ions as a function of pressure. The plotted data and error bars correspond to $\mu_1$, $\mu_2$, $\mu_3$ and $\sigma_1$, $\sigma_2$, $\sigma_3$ in Eq. (4), respectively. The profiles of lattice parameters and ionic displacements as a function of simulation time under the temperature and pressure conditions around the phase transitions are shown in Figs. S4-S9 in Supplemental Materials.

The phase transition pressures were determined by the changes in the relation between the lattice parameters. For example, in a 16×16×16 supercell (Fig. 4(b)), $a = b = c$ at $P \leq 4$ GPa and $P \geq 11$ GPa indicates that the rhombohedral and cubic phases are stable, respectively. Under the intermediate pressures, the orthogonal ($a < b = c$) and the tetragonal ($a = b < c$) phases are stable at $5 \leq P \leq 8$ GPa and $9 \leq P \leq 10$ GPa, respectively. Thus, the R-O, O-T, and T-C phase transition pressures were estimated to be 4-5, 8-9, and 10-11 GPa, respectively. The relative ionic displacements of Ti ions against O ions also exhibit a similar trend to the lattice parameters. The number of nonzero values in the ionic displacements along the a-, b-, and c-axes increases with decreasing pressure. At the high pressures ($P \geq 11$ GPa), all the displacements $\varDelta_a$, $\varDelta_b$, and $\varDelta_c$, are nearly equal to zero in Fig. 5(b), indicating each Ti ion is positioned at the center of the octahedron formed by six O ions in the cubic phase. Similarly, $\varDelta_a \neq \varDelta_b = \varDelta_c = 0$, $\varDelta_a = \varDelta_b \neq \varDelta_c = 0$, and $\varDelta_a = \varDelta_b =$ $\Delta_c \neq 0$ can be seen at $9 \leq P \leq 10$ GPa, $5 \leq P \leq 8$ GPa, and $P \leq 4$ GPa, corresponding to the ionic displacements along [100], [110], and [111] directions in the pseudo-cubic tetragonal, orthogonal, and rhombohedral symmetries, respectively.

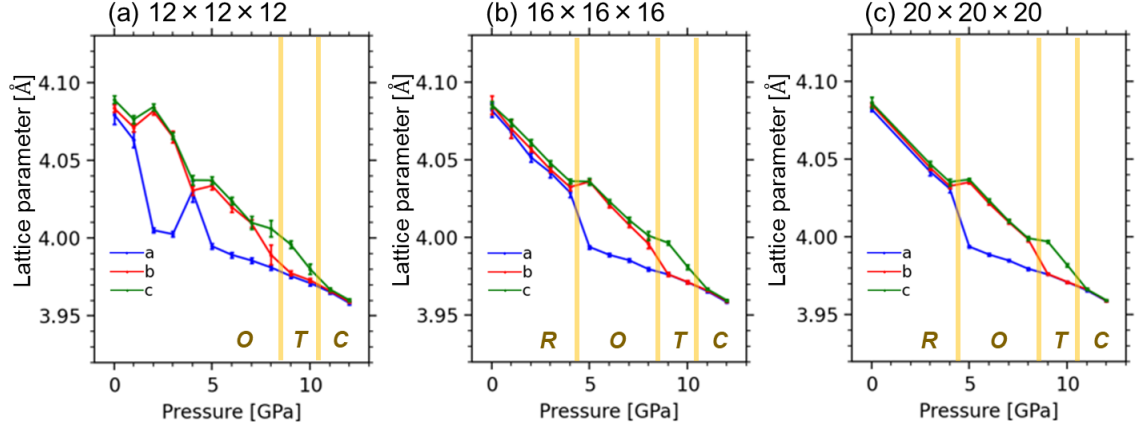

Fig. 4. Lattice parameters as a function of pressure at (a) 12×12×12, (b) 16×16×16, and (c) 20×20×20 supercells. Each lattice parameter was obtained as $\mu_1$, $\mu_2$, or $\mu_3$, with an error bar as $\sigma_1$, $\sigma_2$, or $\sigma_3$ in Eq. (4). The yellow lines correspond to the estimated phase transition pressures, and the letters *R*, *O*, *T*, and *C* correspond to the rhombohedral, orthogonal, tetragonal, and cubic phases, respectively.

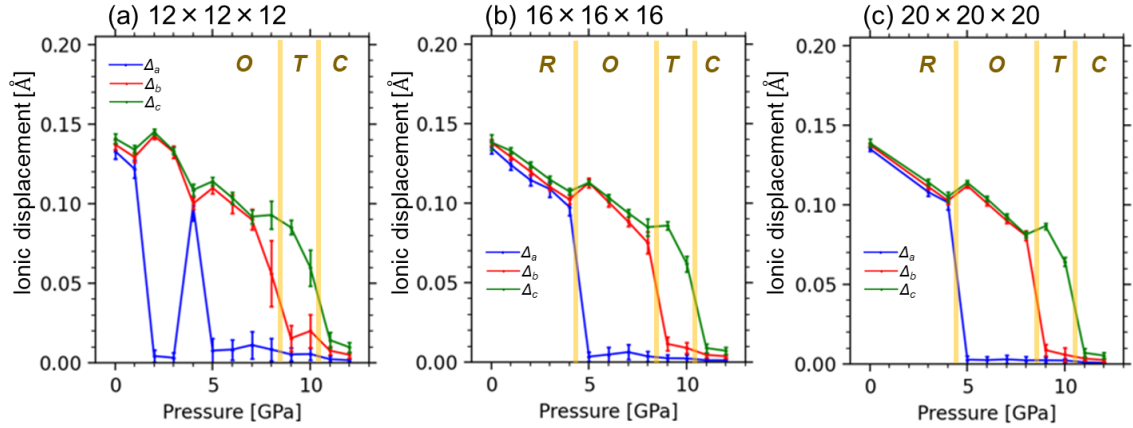

Fig. 5. Relative ionic displacements of Ti ions against O ions along the *a*-, *b*-, and *c*-axes as a function of pressure at (a) 12×12×12, (b) 16×16×16, and (c) 20×20×20 supercells. Each ionic displacement was obtained as $\mu_1$, $\mu_2$, or $\mu_3$, with an error bar as $\sigma_1$, $\sigma_2$, or $\sigma_3$ in Eq. (4). The yellow lines correspond to the estimated phase transition pressures, and the letters *R*, *O*, *T*, and *C* correspond to the rhombohedral, orthogonal, tetragonal, and cubic phases, respectively.

As shown in Figs. 4 and 5, the phase transitions sequentially occur with pressure in the order of *R-O*, *O-T*, and *T-C* transitions in both 16×16×16 and 20×20×20 supercells. By contrast, at the relatively small 12×12×12 cell (Figs. 4(a) and 5(a)), the phase transition pressures cannot be determined clearly. As for the *R-O* phase transition, the lattice parameters and the Ti displacements exhibit the relations of $a = b = c$ and $\varDelta_a = \varDelta_b = \varDelta_c \neq 0$ corresponding to the low-pressure rhombohedral phase at 4 GPa. But the high-pressure orthogonal phase ($a < b = c$ and $\varDelta_a = \varDelta_b \neq \varDelta_c = 0$) appears at 2 and 3 GPa.

The estimated phase transition pressures at 150 K are shown in Table 1. The phase transition pressures in a 16×16×16 supercell coincide with those in a 20×20×20 supercell under this condition. By contrast, the *R-O* phase transition cannot be determined in a 12×12×12 supercell and the *O-T* phase transition pressure is 1 GPa smaller than that in the larger supercells. The cell size therefore should be 16×16×16 or larger. A 20×20×20 supercell was adopted for comparing QTB vs. CTB in this study.

Table 1. Estimated phase transition pressures at 150 K in the QTB-MLPMD simulations, which were simply estimated by averaging the pressures around the phase boundaries in Fig. 4. The *R-O* transition pressure at a 12×12×12 cell was undetermined due to the undesirable result in Fig. 4(a).

|          | *R-O* | *O-T* | *T-C* |
|----------|-------|-------|-------|
| 12×12×12 | Und.  | 7.5   | 10.5  |
| 16×16×16 | 4.5   | 8.5   | 10.5  |
| 20×20×20 | 4.5   | 8.5   | 10.5  |

**B. Comparisons between QTB- and CTB-MLPMD**
 **1. Kinetic energy**

First, the mean kinetic energies were compared between the QTB- and CTB-MLPMD simulations. For reference, the kinetic energies of the lattice vibrations were also estimated under the quantum and classical harmonic approximations (QHA, CHA). The QTB- and CTB-MLPMD were performed under the conditions of $T = 100, 150, 200, 250, 300$ K and $P = 0$ GPa at a 20×20×20 supercell. The kinetic energies of the lattice vibrations under the QHA were obtained by the finite displacement method implemented in phonopy [40,41]. The q-points were sampled at 5×5×5 mesh with the Γ center for the 4×4×4 pseudo-cubic rhombohedral structure after the structural optimization. The threshold for the convergence of the atomic forces in the structural optimization and the amplitude of the finite displacement distance were $1\times10^{-4}$ eV/Å and 0.01 Å, respectively. The kinetic energies of the lattice vibrations under the CHA were estimated to be $1/2 f k_B T$ according to the energy equipartition theory, where *f* corresponds to the DOF of the system.

Figure 6 shows the mean kinetic energies during the QTB- and CTB-MLPMD as a function of temperature. Those of the lattice vibrations under the QHA and CHA are also shown in Fig. 6. The nonzero value at 0 K in the QHA corresponds to the zero-point energy while the mean kinetic energy in the CHA is zero at 0 K. The mean kinetic energies during the CTB-MLPMD depend linearly on the temperature, which coincide with those in the CHA. In contrast, the kinetic energies during the QTB-MLPMD show a good agreement with those in the QHA, in which the temperature dependency becomes smaller at lower temperatures. The mean kinetic energies in the QTB-MLPMD are larger than those in the CTB-MLPMD. This indicates the NQEs are incorporated as the kinetic energy during the QTB-MLPMD, leading to the increase of the mean kinetic energies. The difference in the kinetic energies expands with decreasing temperature, meaning that the NQEs are more significant at lower temperatures.

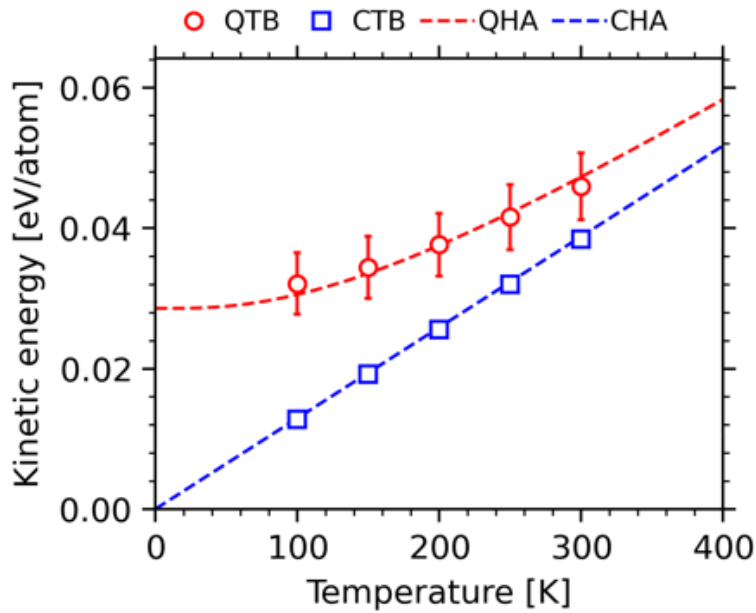

Fig. 6. Mean kinetic energies during the QTB-MLPMD (red circle) and CTB-MLPMD (blue square) with a 20×20×20 pseudo-cubic structure at 0 GPa as a function of temperature, which have an error bar corresponding to a standard deviation. Those of the lattice vibration estimated under the QHA (red dashed line) and CHA (blue dashed line) are also shown.

## 2. Phase diagram

Figure 7 shows the pressure-temperature phase diagrams obtained from the QTB- and CTB-MLPMD simulations, in which the phase boundaries were roughly drawn by linear lines between the estimated phase transition pressures at several temperatures. As in Sec. III.A.3, the phase transition pressures at $T$ = 100, 150, 200, 250, and 300 K were determined by the lattice parameters and the ionic displacements, which are shown as a function of pressure in Figs. S10 and S11 in Supplemental Materials. The estimated phase transition temperatures and pressures in the QTB-MLPMD are lower than those in the CTB-MLPMD. This trend originates from the larger kinetic energy by the NQEs in the QTB-MLPMD as shown in Fig. 6. This corresponds to the increase of apparent temperature in classical statistics, leading to low-temperature and low-pressure shifts of the phase boundaries. In addition, the difference in the phase transition pressures between QTB- and CTB-MLPMD becomes larger with decreasing temperature. For example, the difference in the $T$-$C$ transition pressure is 4 GPa at 100 K vs. 2 GPa at 300 K. This indicates that the NQEs are more significant at lower temperatures and not negligible even at room temperature.

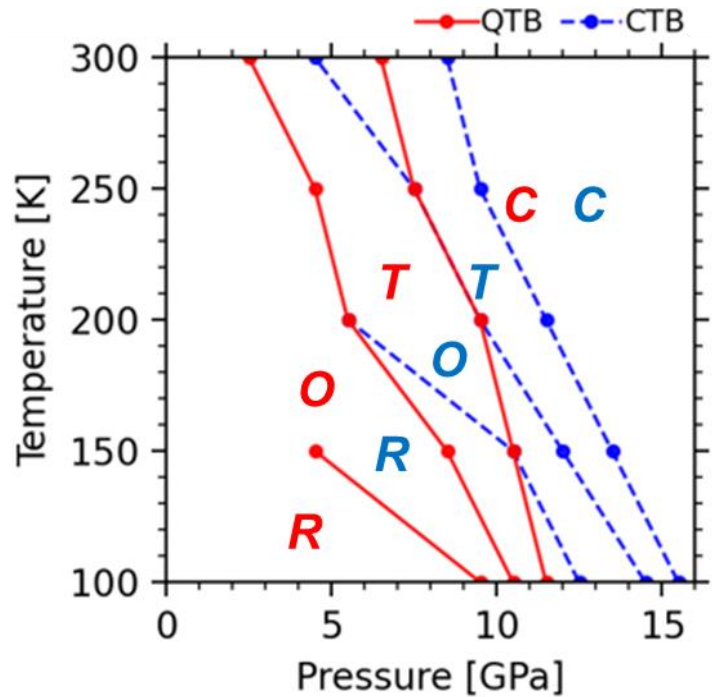

Fig. 7. Pressure-temperature diagrams determined by the QTB-MLPMD (red solid lines) and CTB-MLPMD (blue dashed lines). The letters $R$, $O$, $T$, and $C$ correspond to the rhombohedral, orthogonal, tetragonal, and cubic phases, respectively.

The pressure-temperature phase diagram of the QTB-MLPMD was compared with the phase diagrams reported in the literature of the path-integral MC (PIMC) simulations [14] and the experimental dielectric measurements [10]. Figures 8(a) and (b) show the comparison in the phase diagrams between "the QTB-MLPMD (this study) vs. the dielectric measurements (experiment)" and between "the PIMC vs. the experiment", respectively. Note that the phase diagram of the PIMC was corrected in the previous study. The phase boundaries in [14] were shifted as the *theoretical* zero pressure corresponded to −4.8 GPa [12–14], which was determined by comparing the cell volumes of the four different phases between the theoretical calculations and the experimental measurements. Both the corrected and uncorrected phase diagrams in the QTB-MLPMD and the PIMC are shown in Fig. 8, in which the theoretical zero pressure of the PBE_GGA-based MLPs in the QTB-MLPMD was determined to be 6-7 GPa in a similar manner. Figure S12 in Supplemental Materials shows the cell volumes obtained in the QTB-MLPMD as a function of pressure, with the experimental values at ambient pressure (64.2-64.4 Å$^3$) [42]. The *R-O* phase boundary in the PIMC cannot be seen in Fig. 8(b) because $T_{R-O}$ < 100 K in the literature [14].

Compared with the experimental phase diagram, the phase diagram without the correction in the QTB-MLPMD exhibits the higher phase transition temperatures and pressures, in contrast to the lower values in the uncorrected PIMC. This is mainly due to the difference in the force field, i.e., the effective Hamiltonian based on the LDA in the PIMC simulations vs. the constructed MLPs based on the PBE_GGA in this study. The phase transition temperatures and pressures in BaTiO$_3$ tend to be overestimated (underestimated) by PBE_GGA (LDA), as in the lattice parameters and the cell volumes. With the correction of the theoretical pressure, both the QTB-MLPMD and the PIMC reasonably reproduce the experimental phase diagram, although some discrepancies remain observed.

Comparing the corrected phase diagrams between the QTB-MLPMD and the PIMC, the phase diagram of the QTB-MLPMD shows slightly better agreement with the experimental one than that of the PIMC. This is possibly due to the other computational conditions except the difference between the PBE_GGA and the LDA, such as the differences in the cell size and the DOFs. In the PIMC study, the phase transition temperatures and pressures were determined at a 12×12×12 supercell, in which the DOFs in the system were reduced to the phonon modes corresponding to the TO and acoustic modes of the cubic structure. By contrast, the phase diagram in the QTB-MLPMD was obtained at a 20×20×20 supercell, in which all the DOFs were considered. Note that the *R-O* phase transition pressures were not able to be determined at a 12×12×12 cell with all the DOFs considered in this study, as shown in Figs. 4(a) and 5(a). The larger cell simulations can reproduce the experimental results more accurately in the present study.

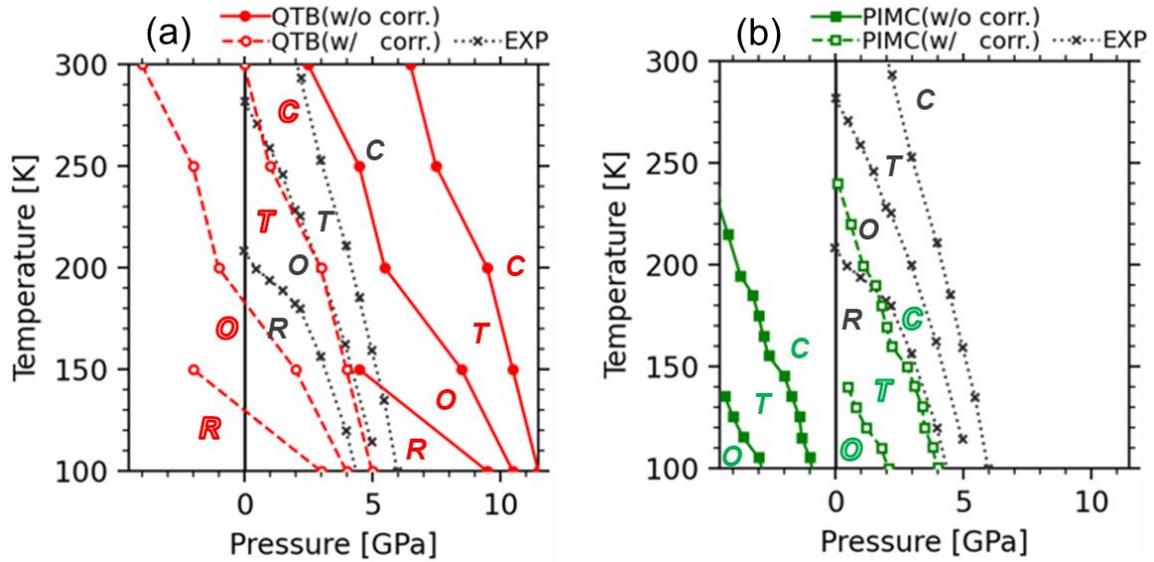

Fig. 8. Comparison of the pressure-temperature diagrams between (a) "the QTB-MLPMD (red lines) vs. the experiment (black dotted lines)", and between (b) "the PIMC (green lines) and the experiment". The phase diagrams without (solid lines) and with (dashed lines) the correction are shown for the QTB-MLPMD and PIMC, in which the theoretical zero pressures are corrected to be +6.5 and −4.8 GPa, respectively. The letters $R$, $O$, $T$, and $C$ correspond to the rhombohedral, orthogonal, tetragonal, and cubic phases. The phase transition temperatures and pressures in the PIMC and the experiment were obtained by scanning the reported phase diagrams in [14] and [10], respectively. The phase boundaries were simply drawn by the linear lines between the obtained data.

**IV. Conclusions**

We have investigated the influence of the NQEs on the pressure-temperature phase diagram in BaTiO$_3$ based on the MD simulations. For incorporating the NQEs into the MD simulations, we employed the QTB method, which was implemented into the VASP code in this study. The QTB method is based on the Langevin MD. In the QTB, the NQEs are incorporated into the system as the kinetic energy through a random force, which is given with time correlations in contrast to a Gaussian random force without time correlations in the conventional CTB. In this study, all the DOFs were taken into consideration, and the MLPs were constructed through the on-the-fly scheme to reduce the computational costs of the MD simulations with a large amount of DOFs.

First, the accuracy of the constructed MLPs for BaTiO$_3$ and the dependencies of the phase transitions on the friction coefficient and cell size were checked. The MLPs test showed that the MLPs can reproduce the first-principles (GGA_PBE) accuracy of the total energy, atomic forces on Ti ions, and stress with an acceptable error. Although systematic errors due to the cell-size difference were seen in the total energy between the MLPs and the first-principles calculations, these errors are expected to be negligible for estimating the phase transition temperatures and pressures. The friction-coefficient test showed that the parameter should be large ($\geq$ 15 THz) for reproducing the phase transitions accurately. The *R-O* transition temperature decreased with decreasing friction coefficient as in the previous study, in which the discrepancy from the experimental value became larger. According to the investigation of the cell-size dependency of the phase transitions, the cell size should be 16×16×16 or larger. The phase transition pressures at a relatively small 12×12×12 cell were smaller than those at larger supercells or were not able to be determined clearly.

Then, we determined the phase diagram as a function of temperature and pressure by the QTB-MLPMD and CTB-MLPMD, quantitatively evaluating the NQEs on the phase transitions. Because of the larger kinetic energy by the NQEs, the low-temperature and low-pressure shifts of the phase boundaries can be seen in the phase diagram of the QTB-MLPMD, compared to that of the CTB-MLPMD. The shifts of the phase boundaries was larger at lower temperatures, corresponding to a larger influence of the NQEs. The obtained phase diagram was finally compared with those reported in the literature, i.e., the PIMC and the experiment. The phase transition temperatures and pressures were larger in the QTB-MLPMD than those in the PIMC, mainly due to the difference in the exchange-correlation functional between GGA_PBE and LDA for constructing the force fields. The phase diagram with the correction of the theoretical zero pressure in the QTB-MLPMD shows good agreement with the experimental results, which was slightly better than that in the PIMC.

**Acknowledgments**

This paper was partially supported by JSPS, KAKENHI (Grant No. 24K01147), JST, PRESTO (Grant No. JPMJPR24J8), and JST, SPRING (Grant No. JPMJSP2110). The supercomputer of ACCMS, Kyoto University was used for FPMD simulations. K.K. was financially supported by the Fujinomori-Masamichi Memorial Scholarship at the Mining and Materials Processing Institute of Japan.

**References**

[1] M. G. Harwood, P. Popper, and D. F. Rushman, Nature **160**, 58 (1947).

[2] H. F. Kay, Acta Cryst. **1**, 229 (1948).

[3] A. von Hippel, Rev. Mod. Phys. **22**, 221 (1950).

[4] B. Ravel, E. A. Stern, R. I. Vedrinskii, and V. Kraizman, Ferroelectrics **206**, 407 (1998).

[5] M. Acosta, N. Novak, V. Rojas, S. Patel, R. Vaish, J. Koruza, G. A. Rossetti, and J. Rödel, Appl. Phys. Rev. **4**, 041305 (2017).

[6] C. Zhao, Y. Huang, and J. Wu, InfoMat. **2**, 1163 (2020).

[7] U. D. Venkateswaran, V. M. Naik, and R. Naik, Phys. Rev. B **58**, 14256 (1998).

[8] G. A. Samara, Phys. Rev. **151**, 378 (1966).

[9] D. L. Decker and Y. X. Zhao, Phys. Rev. B **39**, 2432 (1989).

[10] T. Ishidate, S. Abe, H. Takahashi, and N. Môri, Phys. Rev. Lett. **78**, 2397 (1997).

[11] S. Ehsan, M. Arrigoni, G. K. H. Madsen, P. Blaha, and A. Tröster, Phys. Rev. B **103**, 094108 (2021).

[12] W. Zhong, D. Vanderbilt, and K. M. Rabe, Phys. Rev. Lett. **73**, 1861 (1994).

[13] W. Zhong, D. Vanderbilt, and K. M. Rabe, Phys. Rev. B **52**, 6301 (1995).

[14] J. Íñiguez and D. Vanderbilt, Phys. Rev. Lett. **89**, 115503 (2002).

[15] G. Geneste, H. Dammak, M. Hayoun, and M. Thiercelin, Phys. Rev. B **87**, 014113 (2013).

[16] F. Brieuc, Y. Bronstein, H. Dammak, P. Depondt, F. Finocchi, and M. Hayoun, J. Chem. Theory Comput. **12**, 5688 (2016).

[17] P. Souvatzis, S. Arapan, O. Eriksson, and M. I. Katsnelson, EPL **96**, 66006 (2011).

[18] O. Hellman, I. A. Abrikosov, and S. I. Simak, Phys. Rev. B **84**, 180301 (2011).

[19] I. Errea, M. Calandra, and F. Mauri, Phys. Rev. Lett. **111**, 177002 (2013).

[20] D. M. Ceperley, Rev. Mod. Phys. **67**, 279 (1995).

[21] T. E. Markland and M. Ceriotti, Nat. Rev. Chem. **2**, 3 (2018).

[22] C. P. Herrero and R. Ramírez, J. Phys.: Condens. Matter **26**, 233201 (2014).

[23] H. Dammak, M. Hayoun, F. Brieuc, and G. Geneste, J. Phys.: Conf. Ser. **1136**, 012014 (2018).

[24] H. Dammak, Y. Chalopin, M. Laroche, M. Hayoun, and J.-J. Greffet, Phys. Rev. Lett. **103**, 190601 (2009).

[25] J.-L. Barrat and D. Rodney, J. Stat. Phys. **144**, 679 (2011).

[26] K. Kanayama and K. Toyoura, J. Phys.: Condens. Matter **36**, 445404 (2024).

[27] J. Behler, J. Chem. Phys. **145**, 170901 (2016).

[28] J. Behler and M. Parrinello, Phys. Rev. Lett. **98**, 146401 (2007).

[29] O. T. Unke, S. Chmiela, H. E. Sauceda, M. Gastegger, I. Poltavsky, K. T. Schütt, A. Tkatchenko, and K.-R. Müller, Chem. Rev. **121**, 10142 (2021).

[30] R. Jinnouchi, J. Lahnsteiner, F. Karsai, G. Kresse, and M. Bokdam, Phys. Rev. Lett. **122**, 225701 (2019).

[31] G. S. Grest and K. Kremer, Phys. Rev. A **33**, 3628 (1986).

[32] G. Kresse and J. Hafner, Phys. Rev. B **48**, 13115 (1993).

[33] G. Kresse and J. Furthmüller, Comput. Mater. Sci. **6**, 15 (1996).

[34] G. Kresse and D. Joubert, Phys. Rev. B **59**, 1758 (1999).

[35] P. E. Blöchl, Phys. Rev. B **50**, 17953 (1994).

[36] J. P. Perdew, K. Burke, and M. Ernzerhof, Phys. Rev. Lett. **77**, 3865 (1996).

[37] M. Buchholz, E. Fallacara, F. Gottwald, M. Ceotto, F. Grossmann, and S. D. Ivanov, Chem. Phys. **515**, 231 (2018).

[38] S. Huppert, T. Plé, S. Bonella, P. Depondt, and F. Finocchi, Appl. Sci. **12**, 9 (2022).

[39] E. Mangaud, S. Huppert, T. Plé, P. Depondt, S. Bonella, and F. Finocchi, J. Chem. Theory Comput. **15**, 2863 (2019).

[40] A. Togo, F. Oba, and I. Tanaka, Phys. Rev. B **78**, 134106 (2008).

[41] A. Togo, J. Phys. Soc. Jpn. **92**, 012001 (2023).

[42] G. H. Kwei, A. C. Lawson, S. J. L. Billinge, and S. W. Cheong, J. Phys. Chem. **97**, 2368 (1993).

*Supplemental Materials*

# Quantitative evaluation of nuclear quantum effects on the phase transitions in BaTiO$_3$ using large-scale molecular dynamics simulations based on machine learning potentials

Kansei Kanayama* and Kazuaki Toyoura

*Department of Materials Science and Engineering, Kyoto University, Kyoto 606-8501, Japan*

*kanayama.kansei.67z@st.kyoto-u.ac.jp

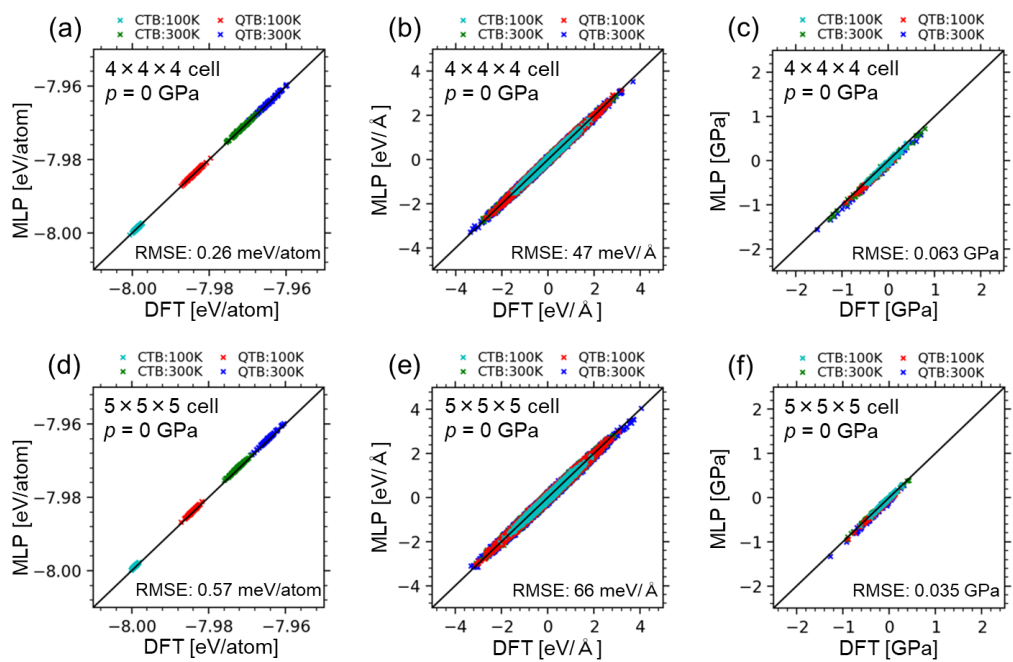

Fig. S1. (a,d) Total energies, (b,e) atomic forces on titanium ions, and (c,f) stresses computed in the constructed MLPs as a function of those in the FP calculations. The data corresponds to the structures sampled during the MLPMD simulations at 0 GPa with (a-c) 4×4×4 and (d-f) 5×5×5 pseudo-cubic structures.

*Supplemental Materials*

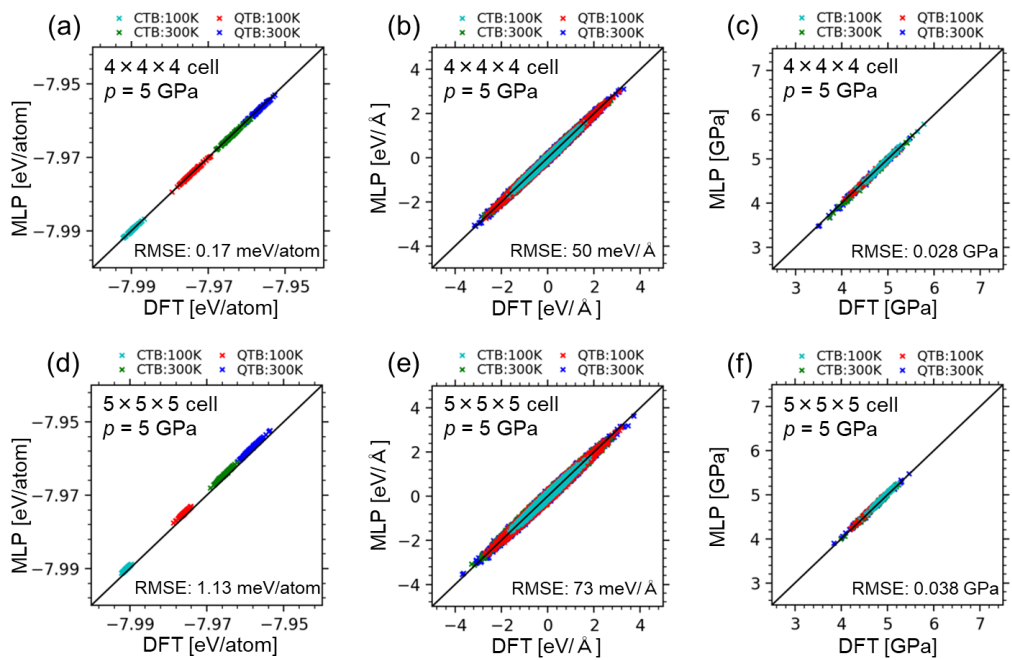

Fig. S2. (a,d) Total energies, (b,e) atomic forces on titanium ions, and (c,f) stresses computed in the constructed MLPs as a function of those in the FP calculations. The data corresponds to the structures sampled during the MLPMD simulations at 5 GPa with (a-c) 4×4×4 and (d-f) 5×5×5 pseudo-cubic structures.

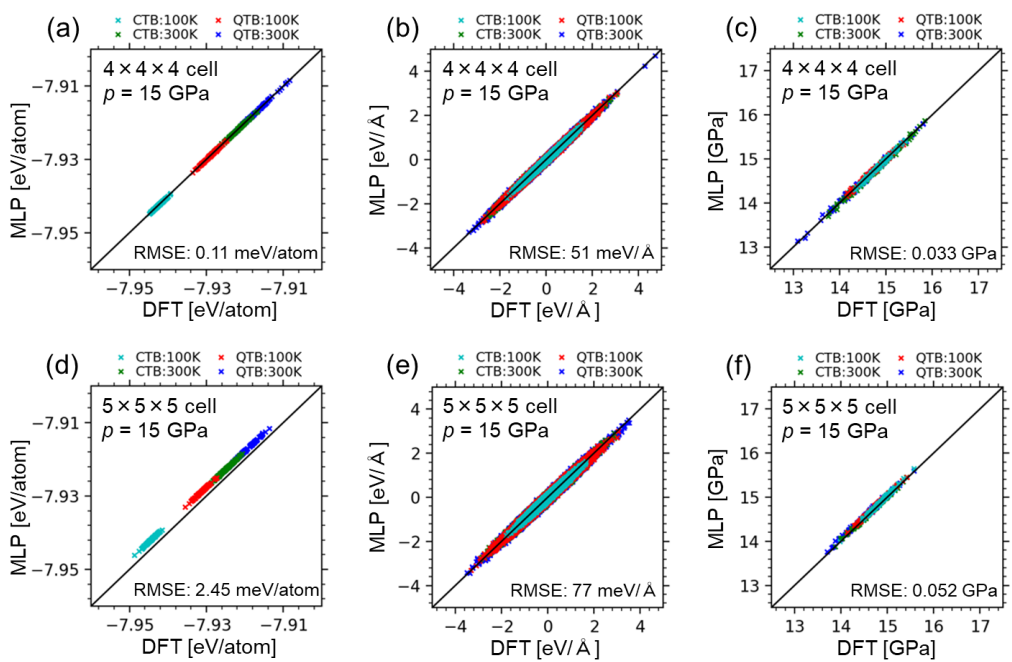

Fig. S3. (a,d) Total energies, (b,e) atomic forces on titanium ions, and (c,f) stresses computed in the constructed MLPs as a function of those in the FP calculations. The data corresponds to the structures sampled during the MLPMD simulations at 15 GPa with (a-c) 4×4×4 and (d-f) 5×5×5 pseudo-cubic structures.

*Supplemental Materials*

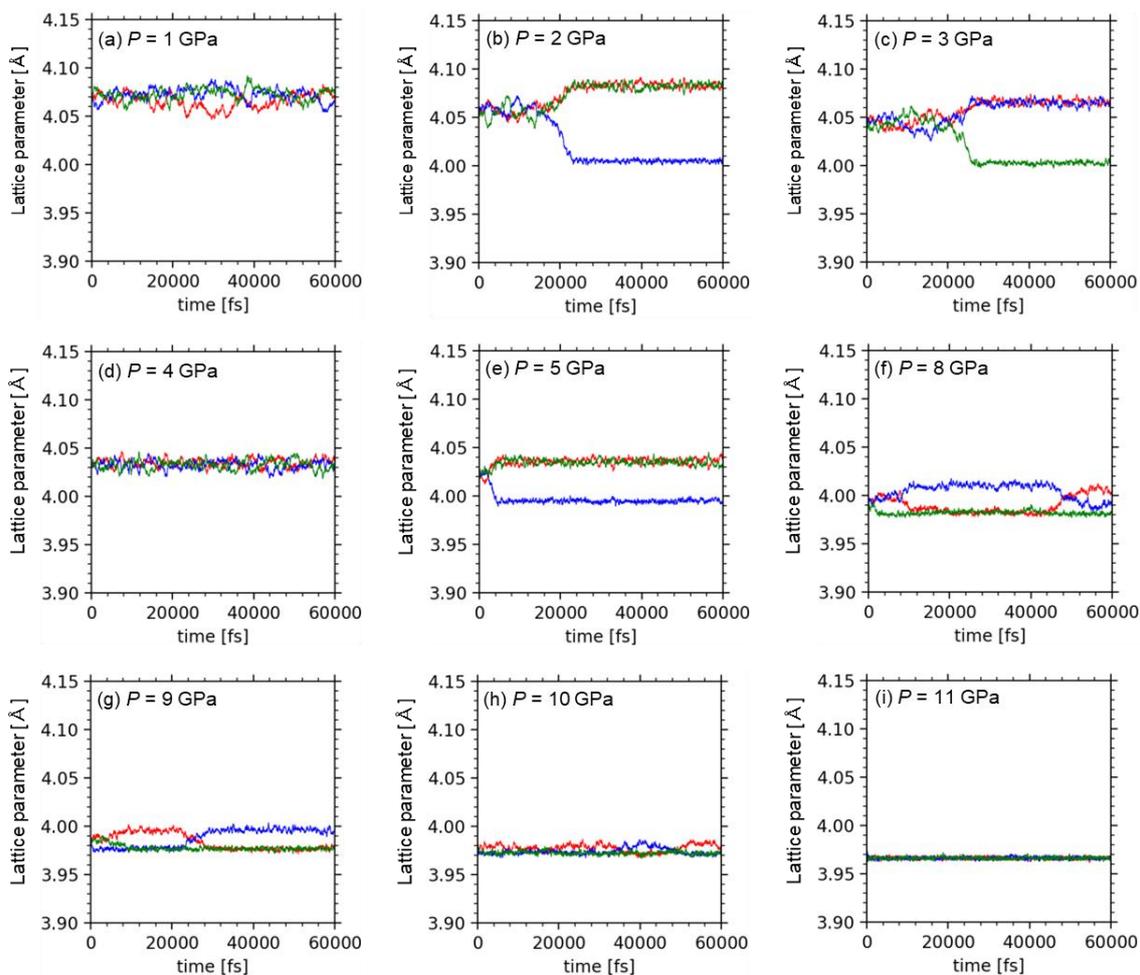

Fig. S4. Lattice parameter profiles as a function of time during the QTB-MLPMD simulations at 150 K in a 12×12×12 pseudo-cubic supercell.

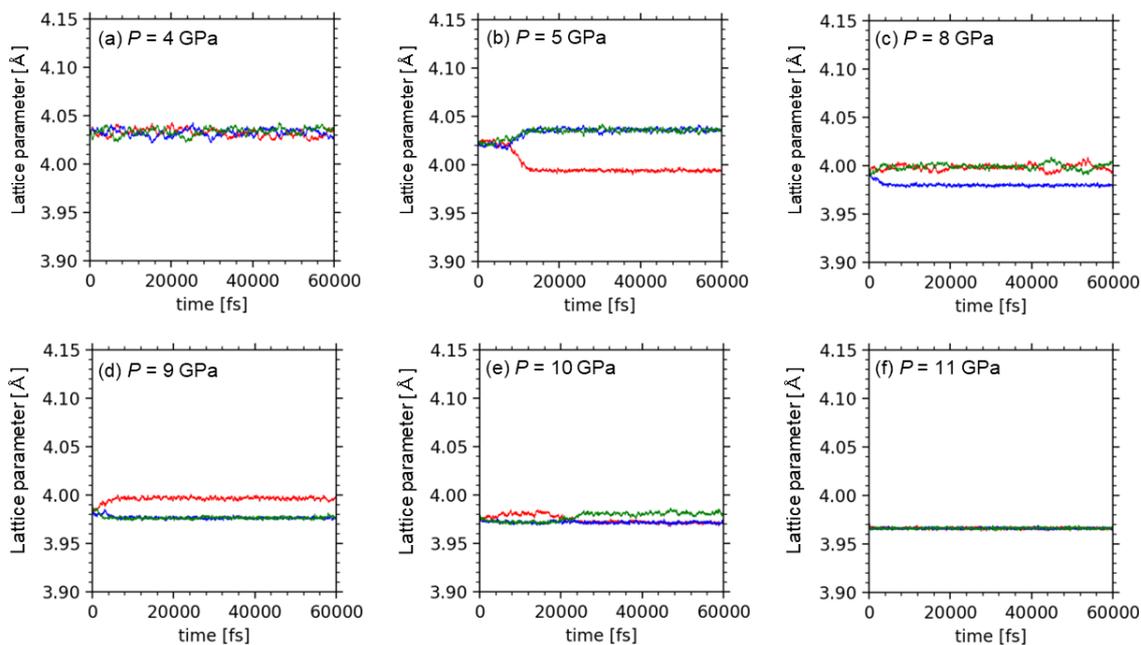

Fig. S5. Lattice parameter profiles as a function of time during the QTB-MLPMD simulations at 150 K in a 16×16×16 pseudo-cubic supercell.

**3** / **8**

*Supplemental Materials*

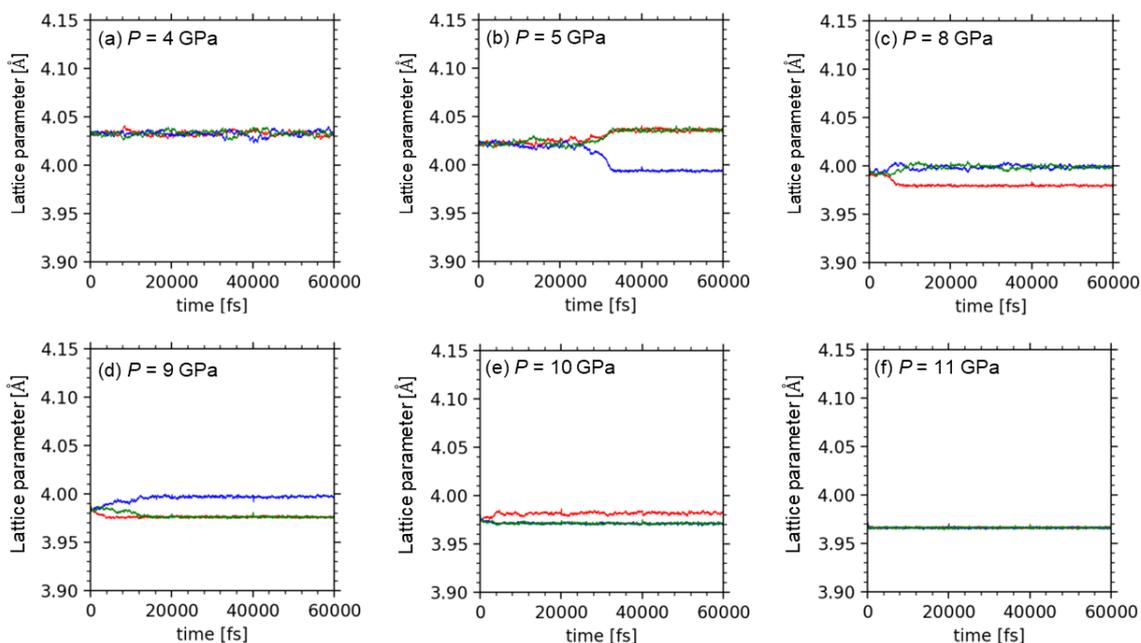

Fig. S6. Lattice parameter profiles as a function of time during the QTB-MLPMD simulations at 150 K in a 20×20×20 pseudo-cubic supercell.

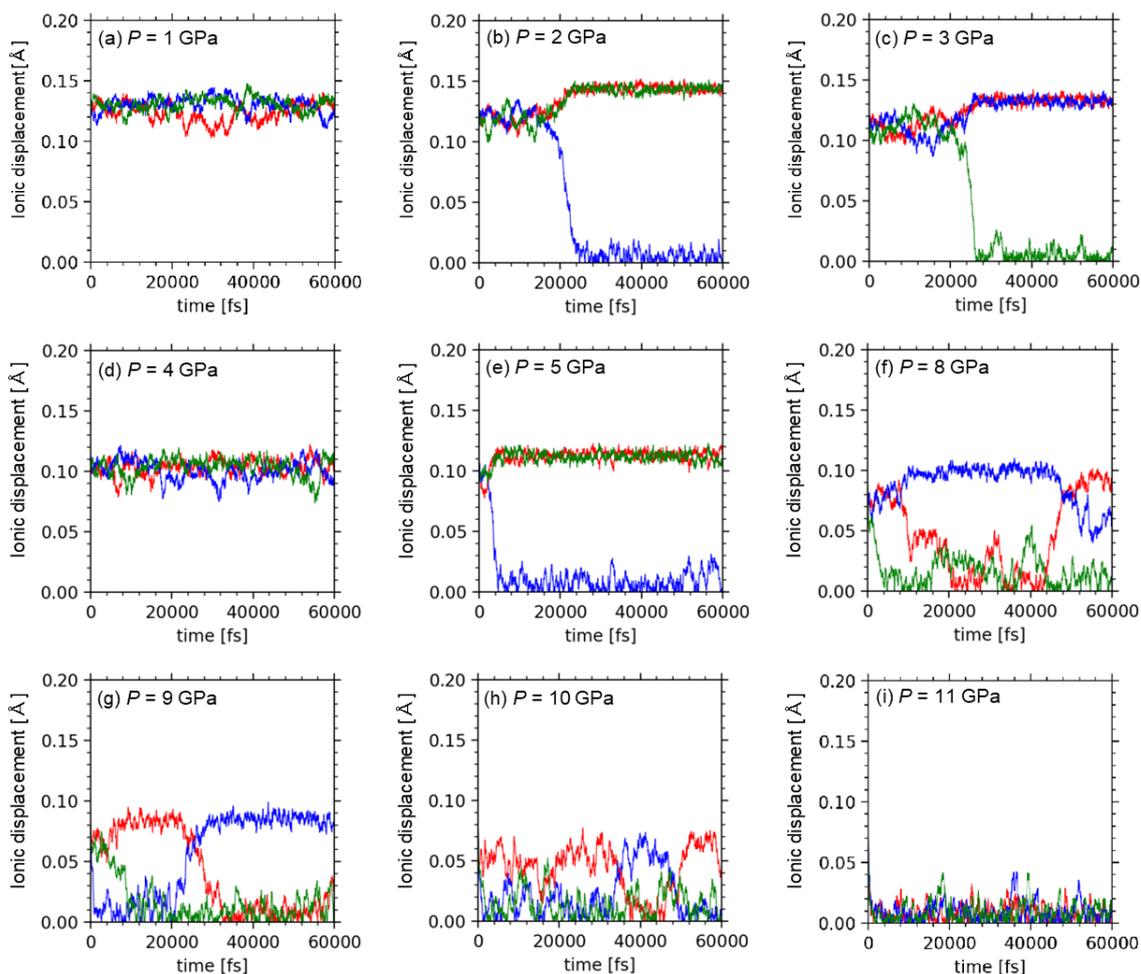

Fig. S7. Profiles of the ionic displacements of Ti ions against O ions along the *a*-, *b*-, and *c*-axes as a function of time during the QTB-MLPMD simulations at 150 K in a 12×12×12 pseudo-cubic supercell.

*Supplemental Materials*

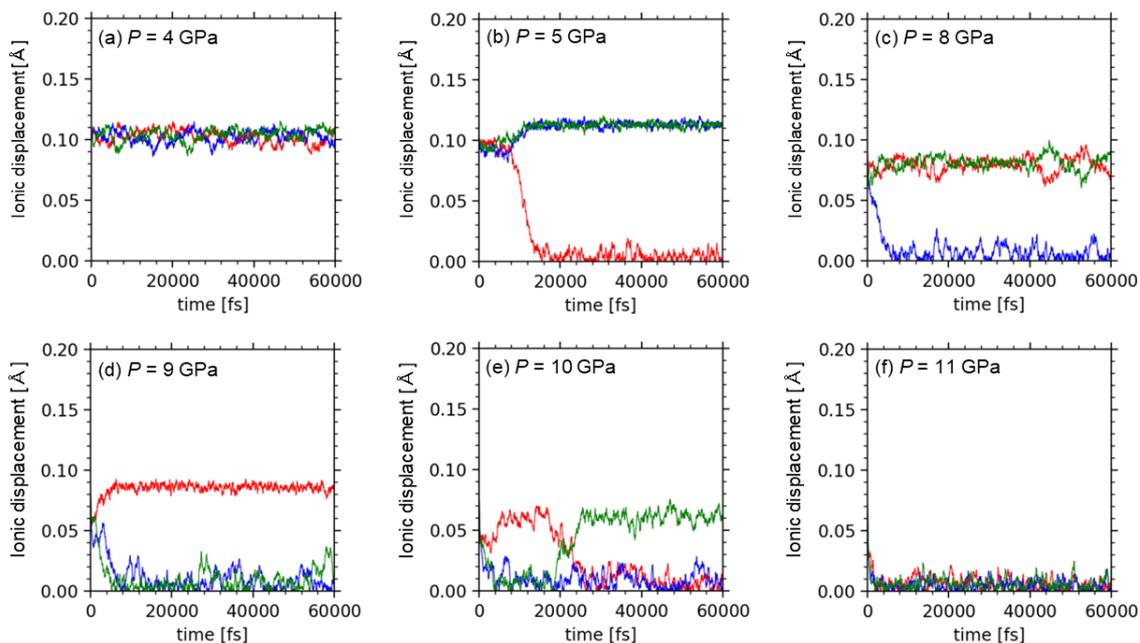

Fig. S8. Profiles of the ionic displacements of Ti ions against O ions along the *a*-, *b*-, and *c*-axes as a function of time during the QTB-MLPMD simulations at 150 K in a 16×16×16 pseudo-cubic supercell.

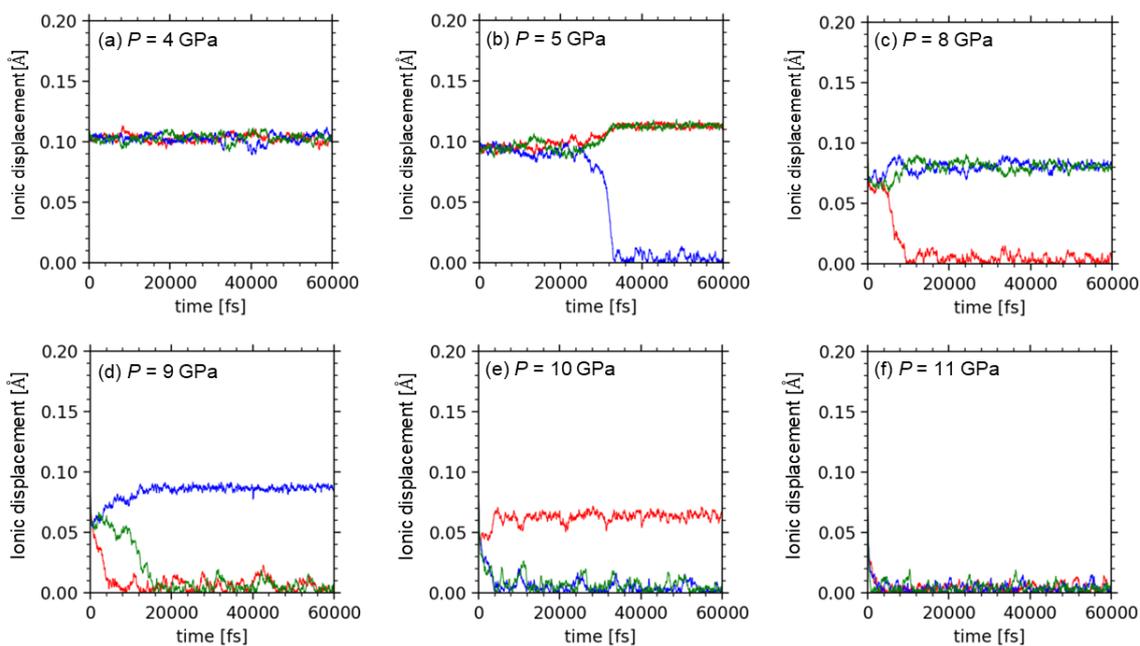

Fig. S9. Profiles of the ionic displacements of Ti ions against O ions along the *a*-, *b*-, and *c*-axes as a function of time during the QTB-MLPMD simulations at 150 K in a 20×20×20 pseudo-cubic supercell.

*Supplemental Materials*

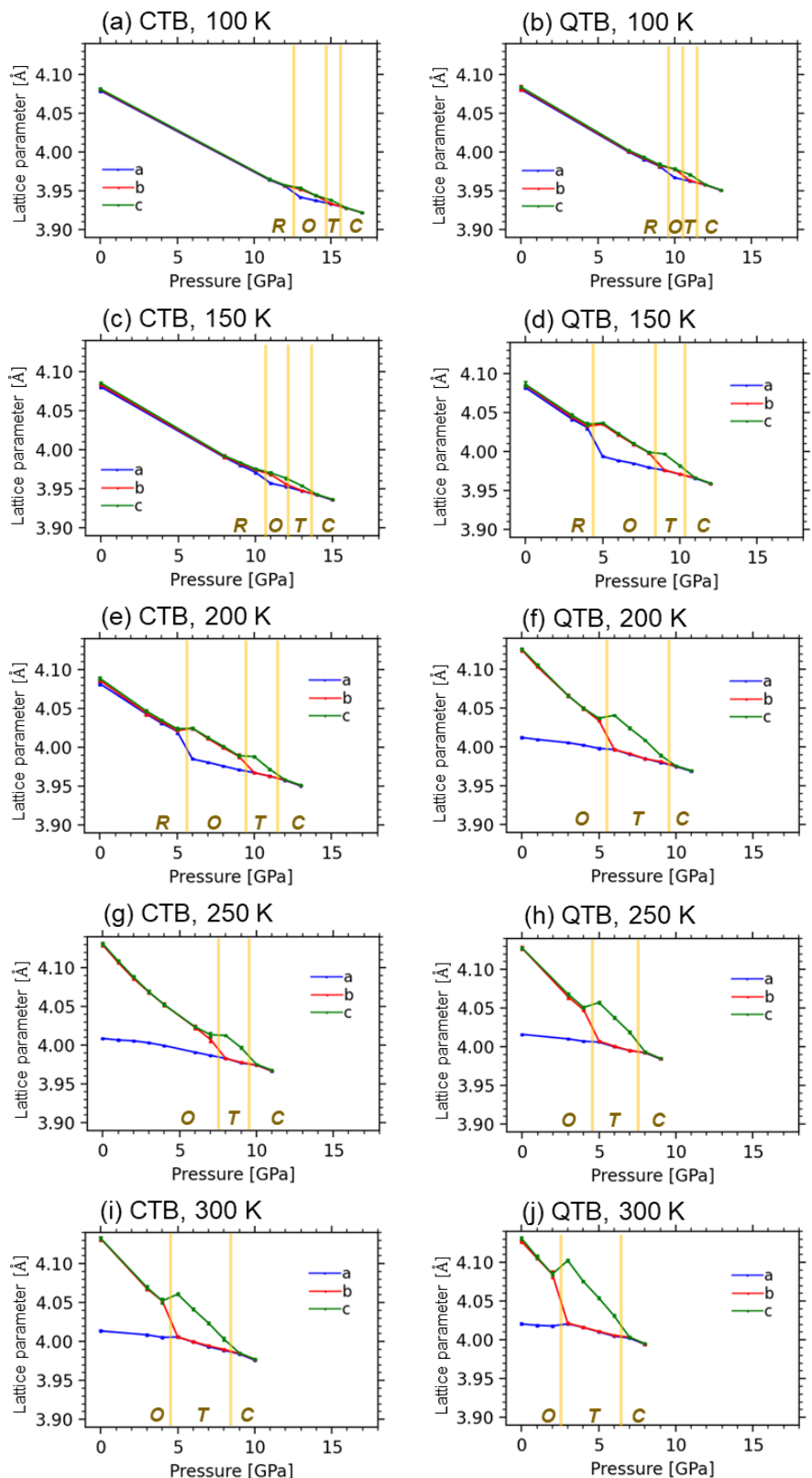

Fig. S10. Lattice parameters as a function of pressure at (a,b) 100 K, (c,d) 150 K, (e,f) 200 K, (g,h) 250 K, and (i,j) 300 K. The MLPMD simulations were performed at a 20×20×20 supercell with the (a,c,e,g,i) QTB and (b,d,f,h,j) CTB, and each lattice parameter was obtained as $\mu_1$, $\mu_2$, $\mu_3$, with an error bar as $\sigma_1$, $\sigma_2$, $\sigma_3$ in Eq. (4). The yellow lines correspond to the estimated phase transition pressures, and the letters *R*, *O*, *T*, and *C* correspond to the rhombohedral, orthogonal, tetragonal, and cubic phases, respectively.

*Supplemental Materials*

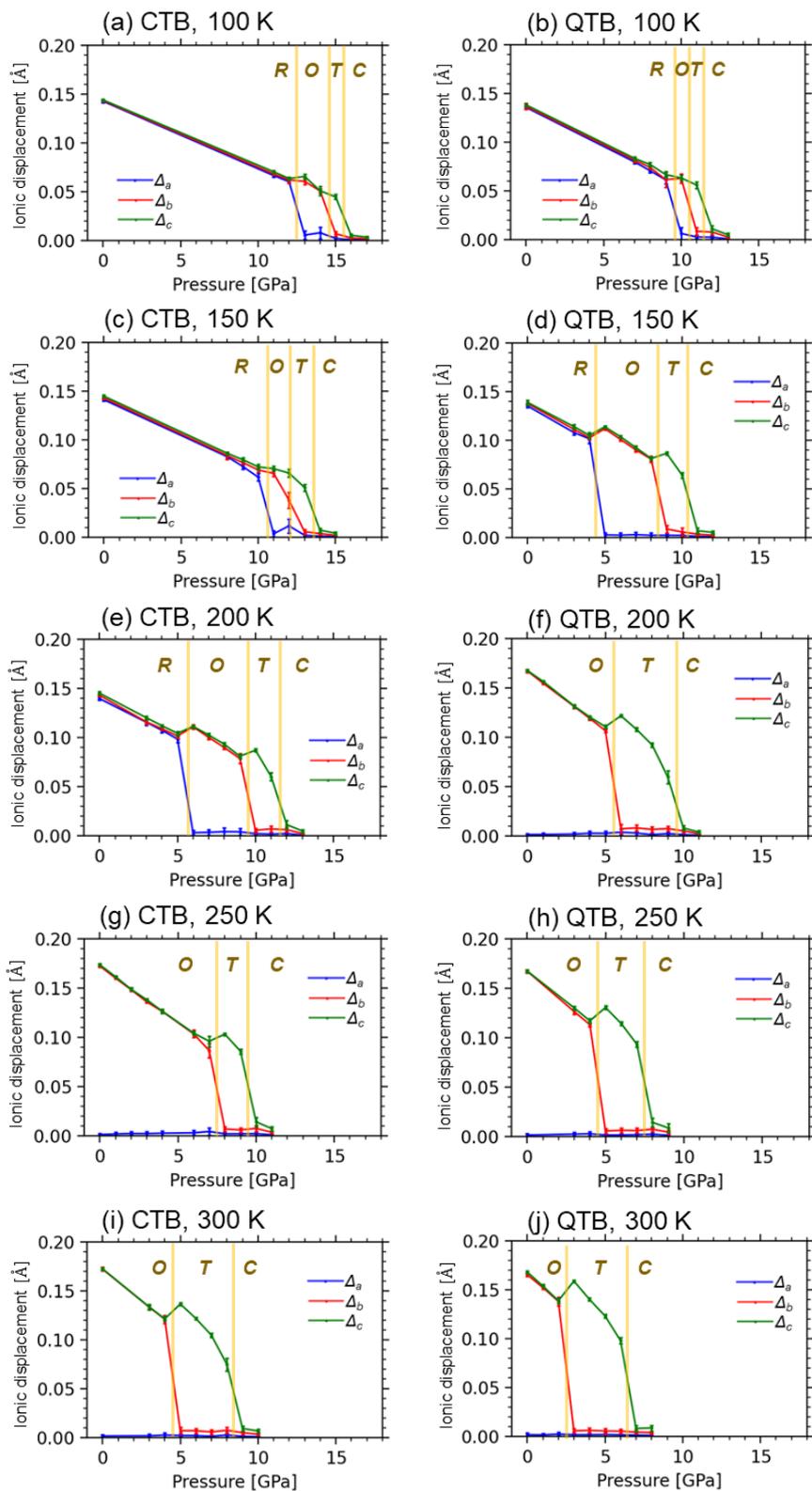

Fig. S11. Relative ionic displacements of Ti ions against O ions along the *a*-, *b*-, and *c*-axes as a function of pressure at (a,b) 100 K, (c,d) 150 K, (e,f) 200 K, (g,h) 250 K, and (i,j) 300 K. The MLPMD simulations were performed at a 20×20×20 supercell with the (a,c,e,g,i) QTB and (b,d,f,h,j) CTB, and each lattice parameter was obtained as $\mu_1$, $\mu_2$, $\mu_3$, with an error bar as $\sigma_1$, $\sigma_2$, $\sigma_3$ in Eq. (4). The yellow lines correspond to the estimated phase transition pressures, and the letters *R*, *O*, *T*, and *C* correspond to the rhombohedral, orthogonal, tetragonal, and cubic phases, respectively.

*Supplemental Materials*

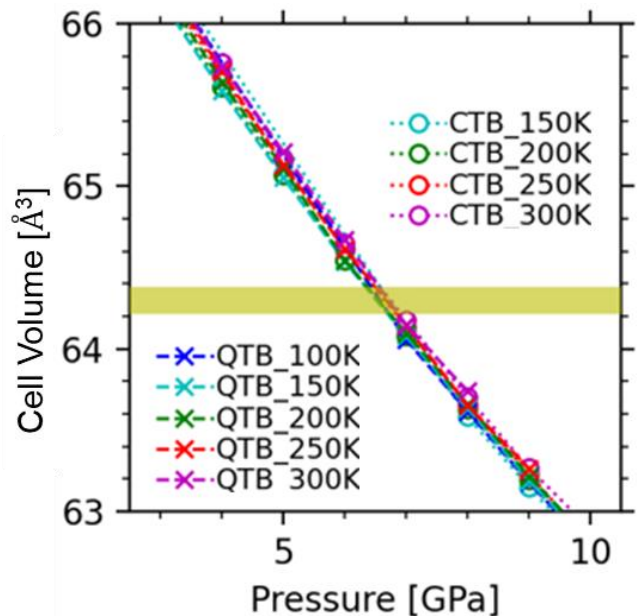

Fig. S12. Cell volume as a function of pressure, which is obtained by averaging the values during the MLPMD simulations after the thermal equilibrium steps. The MLPMD was performed under the conditions of $T$ = 100, 150, 200, 250, and 300 K with the QTB or CTB. The yellow line corresponds to the experimental values (64.2-64.4 Å$^3$) at ambient pressure in the range of 10-350 K.